\documentclass[mathleft,fleqn,%
]{an}
\usepackage[varg]{txfonts}
\usepackage{graphicx}
\usepackage{natbib}
\bibpunct{(}{)}{;}{a}{}{,}
\newcommand{\COBOLD}{{\sf CO$^5$BOLD}}
\newcommand{\CIFIST}{{\sf CIFIST}}

\newcommand{\COROT}{{\sf CoRoT}}
\newcommand{\KEPLER}{{\sf Kepler}}

\newcommand{\pun}[1]{\mbox{\rm\,#1}}

\newcommand{\Teff}{\ensuremath{T_{\mathrm{eff}}}}
\newcommand{\logg}{\ensuremath{\log g}}
\newcommand{\ximic}{\ensuremath{\xi_\mathrm{micro}}}

\newcommand{\beq}{\begin{equation}}
\newcommand{\eeq}{\end{equation}}

\newcommand{\xtmean}[1]{\ensuremath{\left\langle #1\right\rangle}}

\newcommand{\moh}{\ensuremath{\left[\mathrm{M/H}\right]}}

\newcommand{\sig}[1]{{\ensuremath{\sigma_{#1}}}}

\newcommand{\Rstar}{\ensuremath{R}}
\newcommand{\Rsun}{\ensuremath{\mathrm{R}_\odot}}

\newcommand{\sigr}[1]{{\ensuremath{\tilde{\sigma}_{#1}}}}
\newcommand{\siggran}{\ensuremath{\sig{\mathrm{gran}}}}

\newcommand{\vargranr}{\ensuremath{\tilde{\sigma}^2_\mathrm{gran}}}
\newcommand{\siggranr}{\ensuremath{\sigr{\mathrm{gran}}}}

\newcommand{\nuc}{\ensuremath{\nu_\mathrm{gran}}}
\newcommand{\SH}{\ensuremath{S_\mathrm{H}}}

\newcommand{\path}{figures/}

\begin{document}

% The following seven commands are intended for editorial usage and
% should be ignored by the author(s).
\Pagespan{1}{}% Document's page range. 
% If second parameter is left empty, the last page is computed
% automatically.
\Yearpublication{2014}%
\Yearsubmission{2014}%
\Month{0}%   
\Volume{999}%  
\Issue{0}% 
\DOI{asna.201400000}% 

\title{Hydrodynamical model atmospheres: Their impact on stellar
       spectroscopy and asteroseismology of late-type stars}

\author{H.-G.\,Ludwig\inst{1}\fnmsep\thanks{Corresponding author:
        {H.Ludwig@lsw.uni-heidelberg.de}}
% Example for footnote, note the usage of the \texttt{fnmsep} command
% as separator between institute number and footnote mark}
\and M.\,Steffen\inst{2}
}
\titlerunning{Hydrodynamical model atmospheres in spectroscopy and
  asteroseismology}
\authorrunning{H.-G.\,Ludwig \&\ M.\,Steffen}
\institute{Zentrum f\"ur Astronomie der Universit\"at Heidelberg,
  Landessternwarte, K\"onigstuhl 12, 69117 Heidelberg, Germany
\and 
Leibniz-Institut f\"ur Astrophysik Potsdam, An der Sternwarte 16, 
D-14482 Potsdam, Germany
}

\received{XXXX}
\accepted{XXXX}
\publonline{XXXX}


\abstract{%
Hydrodynamical, i.e. multi-dimensional and time-dependent, model atmospheres
of late-type stars have reached a high level of realism. They are commonly
applied in high-fidelity work on stellar abundances but also allow the study
of processes that are not modelled in standard, one-dimensional hydrostatic
model atmospheres. Here, we discuss two observational aspects that emerge from
such processes, the photometric granulation background and the spectroscopic
microturbulence. We use CO5BOLD hydrodynamical model atmospheres to
characterize the total granular brightness fluctuations and characteristic
time scale for FGK stars. Emphasis is put on the diagnostic potential of the
granulation background for constraining the fundamental atmospheric
parameters. We find a clear metallicity dependence of the granulation
background. The comparison between the model predictions and available
observational constraints at solar metallicity shows significant differences,
that need further clarification. Concerning microturbulence, we report on the
derivation of a theoretical calibration based on CO5BOLD models, which shows 
good correspondence with the measurements for stars in the Hyades. We emphasize
the importance of a consistent procedure when determining the microturbulence,
and point to limitations of the commonly applied description of
microturbulence in hydrostatic model atmospheres.}

\maketitle

\section{Introduction}

The structure of late-type stellar atmospheres is mainly shaped by two
processes: radiation and convection. The standard modelling approach of these
atmospheres tries to capture the properties of the stellar radiation field in
great detail while the gas-dynamical aspects are simplified by assuming
one-dimensional (1D), plane-parallel or spherical symmetry. Mixing-length
theory is employed to model the energy transport by convection. In contrast,
hydrodynamical, 3D model atmospheres put emphasis on the representation of
convection, trying to capture the complex geometry and time-dependence of the
gas flows. This comes at the expense of a less detailed treatment of the
radiation field in comparison to 1D models. Nevertheless, the treatment of
radiation is sufficiently accurate to derive the atmospheric structure with a
high degree of realism, e.g., demonstrated by the exquisite reproduction of the
solar center-to-limb variation of the line-blanketed intensity at
different wavelengths
\citep[e.g.,][]{Koesterke+al08,Ludwig+al10}.

The main application of model atmospheres is the analysis of stellar
(electromagnetic) spectra, i.e., the derivation of a star's atmospheric
parameters and its chemical composition. To this end the precise knowledge of
the thermal structure of the stellar atmosphere is essential. And indeed,
differences in the thermal structure between 1D and 3D models cause
differences in the derived parameters and composition. Applications of the 3D
model atmospheres documented in the literature mainly revolve around abundance
issues \citep[see][as recent
  examples]{Amarsi+al15,Caffau+al15,Dobrovolskas+al15,Scott+al15}, and are
commonly rooted in differences between the thermal structure of 1D and 3D
models.

Besides predicting changes in the thermal properties, 3D models provide
information on the flow dynamics taking place in a stellar
atmosphere. Knowledge of the velocity field allows one to eliminate the classical
free parameters of micro- and macro turbulence in spectral analysis. Moreover,
pressure fluctuations associated with the non-stationary flow are believed to
be the exciting agent of solar-like oscillations in late-type stars which
establishes a link to asteroseismology
\citep{Nordlund+Stein01,Samadi+al07b,Samadi+al13a}.
 
In this contribution we intend to discuss ongoing projects in the field of
spectroscopy and asteroseismology besides the main application of 3D models in
abundance work. They are related to i) the diagnostic potential of the
so-called granulation background; and ii) microturbulence calibration 
with 3D models and related insights concerning the deficiencies of classical
1D model atmospheres. The reader should be aware that both examples
are work in progress, so not all issues are fully worked out. 

\section{Granulation background across the Hertzsprung-Russell diagram}

\subsection{The sample of 3D model atmospheres}

\begin{figure}
\includegraphics[width=0.9\columnwidth]{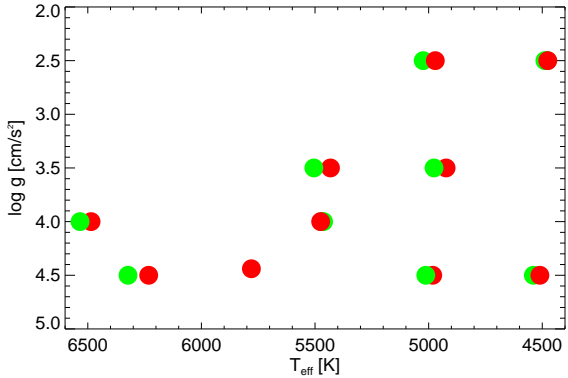}
\caption{The investigated 19 3D models in the \Teff-\logg\ plane. Ten are of
  solar metallicity (red cirles), nine of sub-solar metallicity $\moh=-2$
  (green circles).\label{f:modelsample}}
\end{figure}

Figure~\ref{f:modelsample} depicts the sample of so-called ``local-box'' 3D
models which we used to investigate the granulation background across the
Hertzsprung-Russell diagram (HRD). The models were taken from the \CIFIST\ 3D
model atmosphere grid \citep{Ludwig+al09b} calculated with the
radiation-hydrodynamics code \COBOLD\ \citep{Freytag+al12}, however continued
in time to obtain extra long time series. Ten models assume solar metallicity,
nine models are very metal-poor at $\moh=-2$. The strong metallicity contrast
was chosen to obtain a clear signal related to metallicity effects. The models
cover the main-sequence in the FGK temperature range and the lower red giant
branch, the primary region where \COROT\ and \KEPLER\ targets are located.

\subsection{From local box models to global brightness fluctuations}

A 3D model provides a realization of the convective flows in a representative
volume at the stellar surface. This includes detailed information on the
emergent intensity as a function of surface position, frequency, and
direction. \citet{Ludwig06} laid out a procedure for scaling the radiative
output of a local model to disk-integrated light. Conceptually, the stellar
surface is considered to be tiled by a (potentially great) number of model
``tiles'' or ``patches''. Only the horizontally averaged, frequency-integrated
emergent intensity needs to be considered so that all information taken from a
model is a time series of the bolometric intensity $I_\mathrm{bol}$ at
different limb-angle cosines $\mu_i$, i.e., $I_\mathrm{bol}(t,\mu_i)$ where
$t$ denotes time. The scaling assumes that the horizontal extent of a model
patch is large enough that the emission of neighboring patches is not
correlated. The emission of the ensemble of patches tiling the visible stellar
hemisphere is an outcome of the \textit{incoherent action} of all
patches. This assumption is appropriate for the small-scale random granulation
pattern but inadequate for oscillatory modes which are coherent on the global
stellar scale. In 3D models acoustic oscillations take place as well, however,
they manifest themselves as eigenmodes of the computational domain, and their
amplitudes cannot be directly compared with observed modes.

One does not need to perform the tiling procedure explicitely. The relevant
property of each model tile is its surface area fraction contributing to the
total flux. For each parameter combination the simulations provide only a
single time series of the emergent intensities. To generate other
statistically independent realizations for other tiles the phases of the
Fourier components of the simulated time series are shifted randomly and
independently \citep[see][for details] {Ludwig06}.

The outcome of the scaling procedure is an estimate of the power spectrum of
observable, global brightness fluctuations of a star related to
granulation. An input parameter to the scaling that was not mentioned yet is
the stellar radius $\Rstar$. $\Rstar$ is not a control parameter of a local 3D
model nor a result of the calculation. It is an external piece of information,
and needs to be provided by other consideration in the scaling procedure,
e.g., by considering evolutionary stellar models. The total
(frequency-integrated) granulation-related brightness fluctuations $\siggran$
scale inversely proportional to $\Rstar$. To avoid ambiguities and
uncertainties stemming from the choice of $\Rstar$ we usually work with a
scaled measure of the brightness fluctuations $\siggranr$ defined as
\beq
\siggranr \equiv \frac{\Rstar}{\Rsun}\,\siggran
\eeq
where \Rsun\ denotes the solar radius. \siggranr\ can be calculated from the
local box models alone.

The duration of a time series of 3D models is limited by the affordable
computing time. In comparison to observational time series the simulated series
are short, making the resulting power spectra noisy. To extract the
granular signal from the calculated power spectrum and eliminate the
contribution of oscillatory modes we fit a simple analytical model to the
simulated spectrum of the form
\beq 
\frac{dP}{d\nu}(\nu) = \frac{\vargranr}{\nuc} \exp\left(-\nu/\nuc\right)
     + \mbox{sum of Lorentzians}\,.
\label{e:fitting}
\eeq
\nuc\ is the characteristic granular frequency. A sum of Lorentzian-shaped
peaks model the contribution of the (usually two or three) oscillatory modes
present in the time series. As final result a 3D model provides a prediction
of the scaled granular brightness fluctuations \siggranr\ and characteristic
frequency \nuc. We later use this to construct an ``inverse'' HRD of
convective properties (see Sect.~\ref{s:invhrd}).

\begin{figure}
\includegraphics[width=\columnwidth]{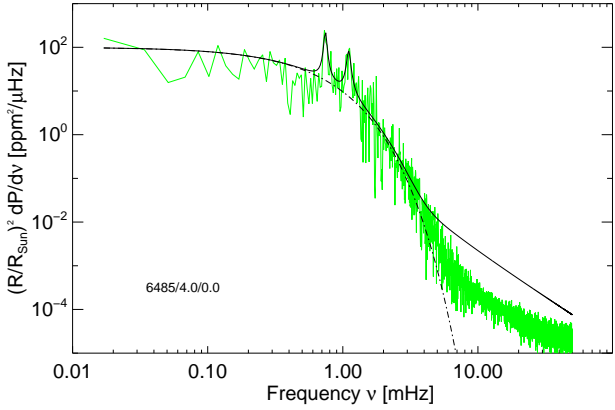}
\caption{Example of a synthetic power spectrum for an F-dwarf model (green
  solid line), and fitted power spectra including (black solid line) and
  leaving-out (dashed-dotted line) p-modes. For display purposes the synthetic
  power spectrum was smoothed
  \label{f:powerspectrum}}
\end{figure}

Figure~\ref{f:powerspectrum} shows %an example of 
a synthetic power spectrum
together with a fit according to Eq.\,(\ref{e:fitting}). The high frequency part
($>6\pun{mHz}$) was not included in the fitting and should be ignored. The
figure depicts the typical run of the granulation background signal in the
power spectra. The frequencies of the (two) box modes roughly coincide with the
mode frequencies expected in a star since the frequencies where modes
are excited are governed by surface convection.

\subsection{``Inverse'' Hertzsprung-Russell diagram of convective properties 
             \label{s:invhrd}}

\begin{figure}
\includegraphics[width=\columnwidth]{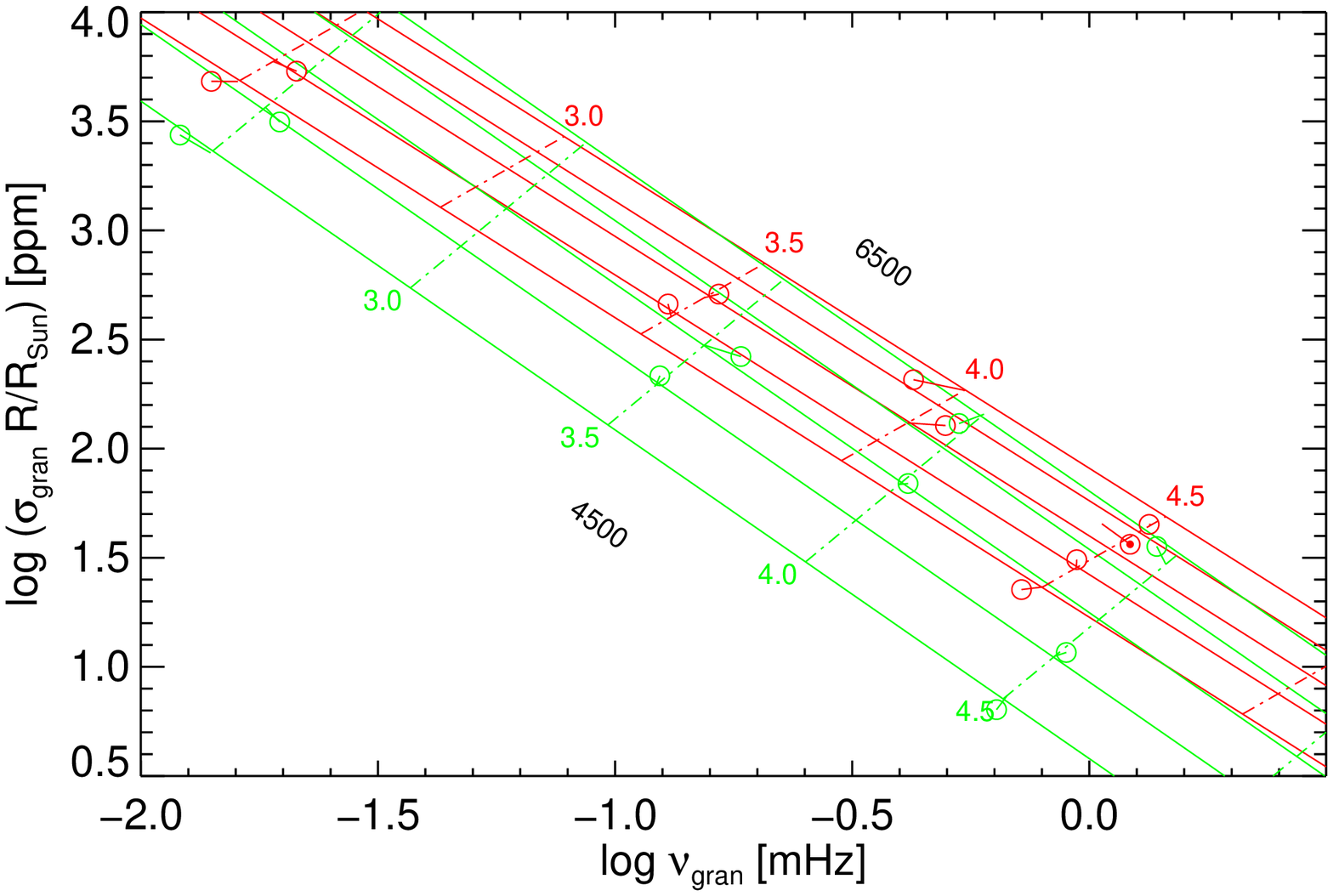}
\caption{Scaled granular brightness fluctuations
  $\siggranr=\Rstar/\Rsun\, \siggran$ and characteristic granular
  frequencies \nuc\ of the 3D models. Red circles mark the location of models of
  solar metallicity, green circles of $\moh=-2$. The lines depict bilinear
  fits to the data points labelled by (log) surface gravity (dashed-dotted
  lines), and effective temperature (solid lines) running from 4500\pun{K} to
  6500\pun{K} in steps of 500\pun{K}. Short line segments attached to the data
  points depict the fitting residuals.\label{f:convectivehrd}}
\end{figure}

Figure~\ref{f:convectivehrd} depicts the derived \siggranr\ and \nuc\ in the
HRD. Instead of plotting the values as function of \Teff, \logg, and \moh\ we
provide the inverse functions to emphasize the diagnostic potential of the
granulation-related properties. Measurements of \siggran\ and \nuc\ from
observed time series, together with information on the stellar radius and
metallicity allow us to constrain the fundamental atmospheric parameters
\Teff\ and \logg. In Fig.~\ref{f:convectivehrd} bilinear fits for both
metallicities are shown separately. A simultaneous fit of both metallicity sets
did not work satisfactorily, presumably due to the large jump in metallicity
between them. The bilinear fits were performed in log-space corresponding to
power laws in linear space commonly used in asteroseimic scaling
relations. The fitting residuals indicate that there is perhaps structure
beyond simple power laws, however, the estimated uncertainties of the model
results (due to limited length of the time series) do not allow to make a
definite statement.

\subsection{Discussion}

The most robust result of our investigation is the about twice as high
temperature sensitivity of the brightness fluctuations at $\moh=-2$ in
comparison to the sensitivity at solar metallicity (as indicated by the widths
of the bands in Fig.~\ref{f:convectivehrd}). However, the magnitude of the
sensitivity -- at least at solar metallicity -- is apparently not compatible
with observations: \citet{Kallinger+al14} observe a smaller T-sensitivity than
predicted by \citet{Samadi+al13b}. Samadi and collaborators also use
\CIFIST\ 3D models for deriving the brightness fluctuations but applying a
different methodology as we do here. Nevertheless, our T-sensitivities are
compatible with the values derived by Samadi et al. so that we arrive at the
same conclusion as Kallinger et al.. The reason for the discrepancy is
presently unclear but further investigations are under way.

\citet{Bastien+al13} and \citet{Cranmer+al14} promote the so-called ``8-hour
flicker'' as gravimeter in late-type stars. We filtered our simulated light
curves similar to the procedure when measuring the 8-hour flicker, and
compared the resulting brightness fluctuations with observational data
presented by \citet{Cranmer+al14}. We applied a bolometric correction as given
by \citet{Ballot+al11} to the observed fluctuations. We find (not shown) a
reasonable agreement in giants. However, in dwarfs the 3D models predict a
weaker 8-hour flicker than observed. There are various factors that may play a
role: The observational data still contains the oscillatory component
enhancing the signal which is absent in the theoretical prediction. Moreover,
stellar activity which is not included in out 3D models might enhance
fluctuations. In contrast, granulation in hot (F-type) stars should suffer
from magnetic suppression \citep{Cranmer+al14,Ludwig+al09a} reducing the
observed signal. Which effect dominates, or whether the cause lies in the
mentioned aspects at all, is not clear yet. However, the 8-hour flicker is not
very-well adapted to dwarfs since it measures only a small fraction of the
underlying granular signal due to the rather long time scales the 8-hour
signal puts emphasis on. Hence, we consider the discrepancy in the 8-hour
flicker not particularly problematic.

All in all one is confronted with the situation that the correspondence
between model prediction and observations is not satisfactory. We expect that
a fully consistent analysis of observational and model data (e.g., applying
the same granular background model) will lead to improvements. We do not think
that a refined treatment of the radiative transfer will help since the bulk of
the radiation emerges from the low photosphere which is already represented
well. Beyond that, we think it becomes apparent that some physics is missing
in the models where magnetic activity related effects are prime suspects
here. It will be interesting to identify more clearly where in the HRD this is
of importance.

\section{Microturbulence issues}

In 1D abundance analyses, the microturbulence~\ximic\ is usually considered a
``nuisance'' parameter since it needs to be determined without being of
primary interest. The microturbulence influences the strength of partly
saturated lines and is interpreted as the effect of the small-scale photospheric
velocity field that 1D models are not accounting for. Less evident, the
microturbulence can also compensate offsets in the thermal structure between
model and star. It is usually described by a depth-independent, isotropic  
Gaussian broadening of fixed width \ximic, adjusted to make weak and strong 
lines provide the same abundance. At low spectral resolution or low 
signal-to-noise ratio -- a typical situation encountered in large scale 
spectroscopic surveys -- abundance analysis relies on strong lines. Strong 
lines are not optimal for abundance work but under the given circumstances 
the only available abundance indicators. If only strong lines are detectable, 
no measurement of \ximic\ is possible so that one is in need of an either 
empirical or theoretical calibration giving the microturbulence as function of
effective temperature, surface gravity, and metallicity. 3D models predict
atmospheric velocity fields, and can provide theoretical estimates. It is
perhaps fair to say that the observational picture of the microturbulence is
rather messy, in the sense that the existence of a tight correlation between 
\ximic\ and stellar spectral type seems not evident. In part, this may be 
related to commonly neglected processes when discussing the microturbulence 
like stellar rotation or activity. Another problematic aspect lies in the fact 
that the derived value of the microturbulence also depends on the set of 
lines selected for its determination.

\begin{figure}
\begin{center}
\mbox{\includegraphics[width=0.65\columnwidth]{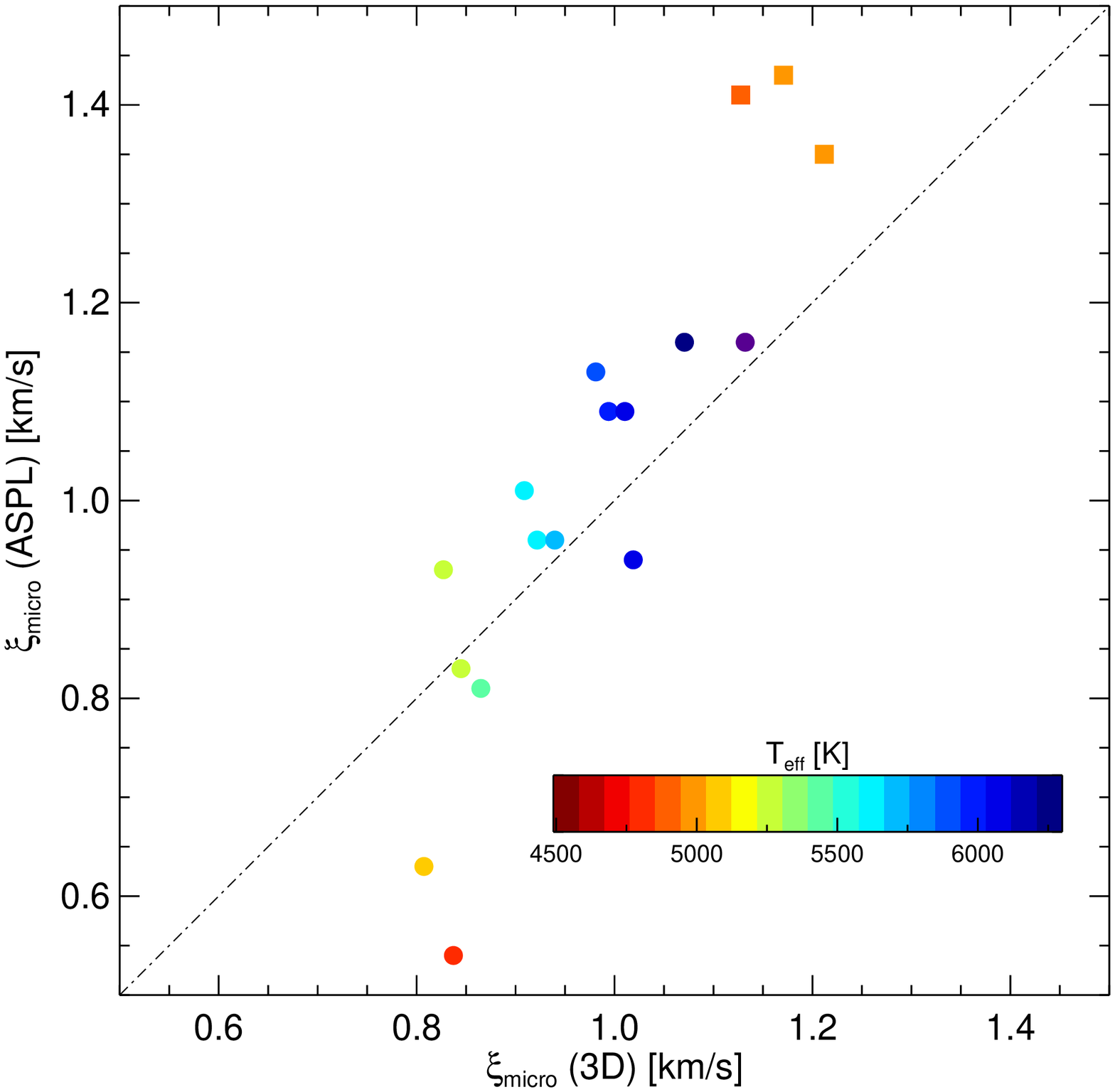}}
\mbox{\includegraphics[width=0.65\columnwidth]{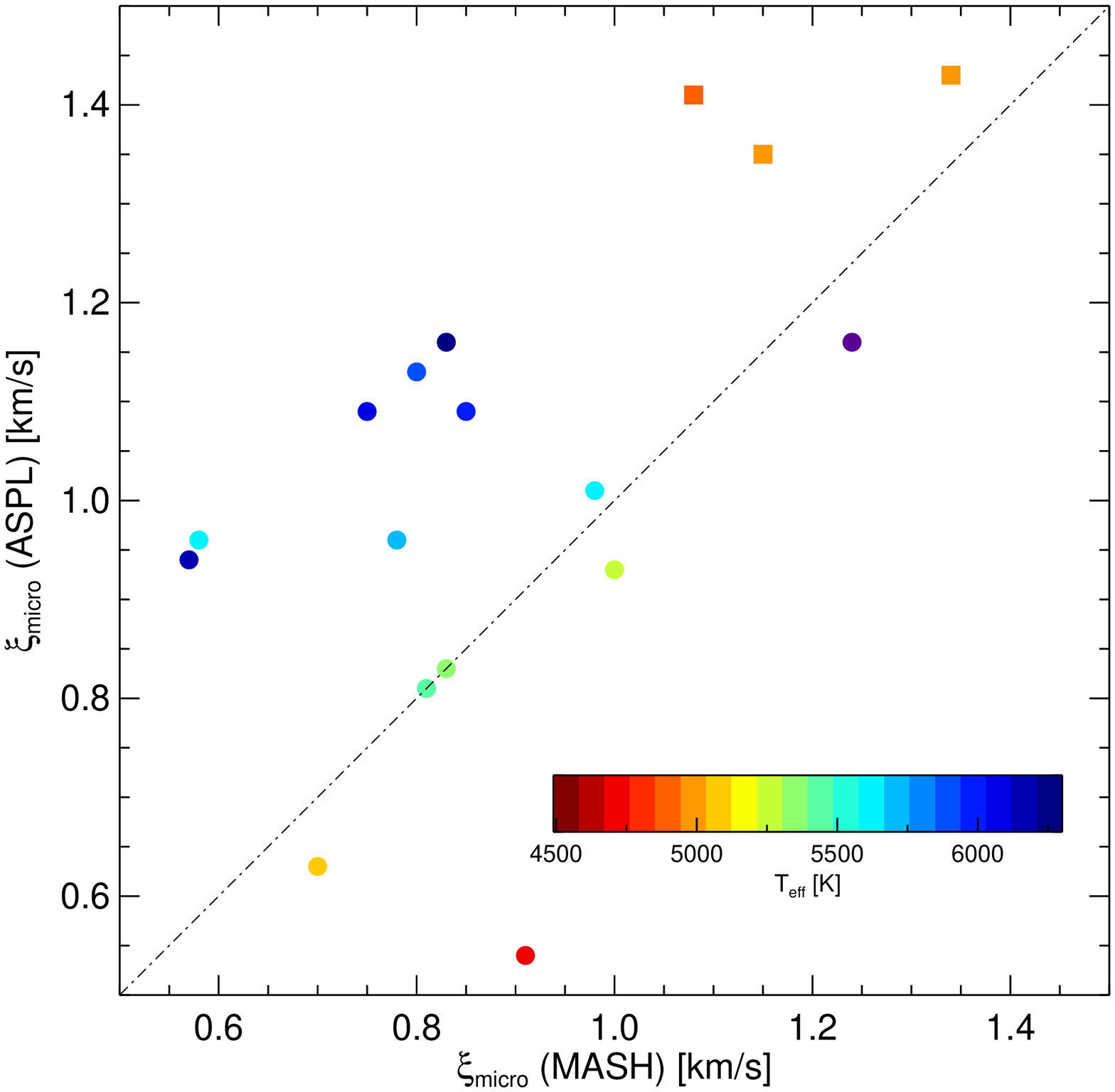}}
\end{center}
\vspace*{-3mm}
\caption{\emph{Top}: Microturbulence of dwarfs (circles) and giants (squares) 
  in the Hyades. Measurements based on \ion{Fe}{i} lines taken from the 
  ``ASPL'' list are plotted against the 3D predicted values. Color encodes 
  stellar effective temperature. The dashed-dotted line indicates the 
  one-to-one correspondence. \emph{Bottom}: Comparison of the empirical 
  microturbulence measurements based on, respectively, the ``ASPL'' and 
 ``MASH''  iron line list.\label{f:aspl3d}}
\end{figure}

\subsection{Microturbulence from 3D models}

3D models can provide reasonable predictions of the microturbulence necessary
for 1D analysis work. \citet{Steffen+al13} laid out different procedures of
how to derive \ximic\ from 3D models. \citet{Dutra+al15} give a recent example
of applying a calibration of \ximic\ from 3D models. They studied dwarfs and 
giants in the Hyades using high-quality UVES and HARPS spectra. Besides using 
their calibration based on 3D models, they also determined the microturbulence 
in the classical spectroscopic way. For this purpose, they applied two
different lists of iron lines, named ``ASPL'' and ``MASH''. 
Figure~\ref{f:aspl3d} (top) shows the comparison between the 3D calibrated 
microturbulence and the empirical values obtained with the ``ASPL'' line list. 
For each star, the 3D microturbulence was obtained from the analytical
expression by Dutra-Ferreira et al. (their Eq.\,2) with the spectroscopic
values of \Teff\ and \logg\ given in their Table\,6. Note that the 3D 
\ximic\ calibration is based on the same selection of ``ASPL'' lines as
used for the classical spectroscopic microturbulence determination.

The correspondence is quite satisfactory for stars with
\Teff\ above $5000$\,K, where deviations are typically less than 0.1\pun{km/s}.
For the two cool dwarfs, the microturbulence predicted by the 3D models seems 
significantly higher than inferred empirically, but we note that the 
measurement uncertainty is rather high for these objects, about
0.2\pun{km/s}. For the three giants, we find the opposite situation. Here the 
microturbulence predicted by the 3D models is too low. This may be attributed 
to an insufficient spatial resolution of the 3D simulations for the giants 
\citep[see][for a discussion of the resolution dependence of \ximic\
derived from the 3D models]{Steffen+al13}. 

Dutra-Ferreira et al. also derived the microturbulence of their target stars
based on the MASH line list. Figure~\ref{f:aspl3d} (bottom) compares the 
resulting values between the two lists, emphasizing the point that the derived
microturbulence depends on the chosen line list (systematic offset, sizable 
scatter). 
%The ASPL based values are systematically larger than MASH based values, 
%and there is also a sizable scatter. 
One should appreciate that ambiguities in the microturbulence 
determination translate into systematic uncertainties in atmospheric
parameters and chemical abundances.

%\begin{figure}
%\begin{center}
%\includegraphics[width=0.7\columnwidth]{\path/asplmash.eps}
%\end{center}
%\caption{Like Fig.~\ref{f:aspl3d}, but comparing the empirical 
%microturbulence measurements based on the ``ASPL'' and ``MASH'' 
%iron line lists.
%\label{f:asplmash}}
%\end{figure}

\subsection{Center-to-limb trouble}

It has been known for a long time from the analysis of spectra taken at 
different positions across the solar disk that microturbulence is a function
of the limb angle $\theta$ ($\mu$\,=\,$\cos\theta$\,=\,$1$ at disk center, 
$\mu$\,=\,$0$ at the limb). According to
\citet{HGR1978}, \ximic\ $\approx 1.0$\,\pun{km/s} at disk center, while
\ximic\ $\approx 1.6$\,\pun{km/s} near the solar limb. Our 3D hydrodynamical 
solar \COBOLD\ model predicts a quantitatively very similar behavior, in 
which the detailed variation with $\mu$ depends
on the properties of the spectral line under consideration. This is a clear
indication that the usual assumption of a depth-independent, isotropic 
microturbulence represented by a single parameter \ximic\ (see previous 
section) provides a poor description of the properties of the the small 
scale velocity field in the solar photosphere. Potentially, this means that 
any model assuming an isotropic microturbulence is prone to systematic 
uncertainties.

\begin{figure}
\mbox{\includegraphics*[bb=242 0 721 205,width=0.95\columnwidth]
{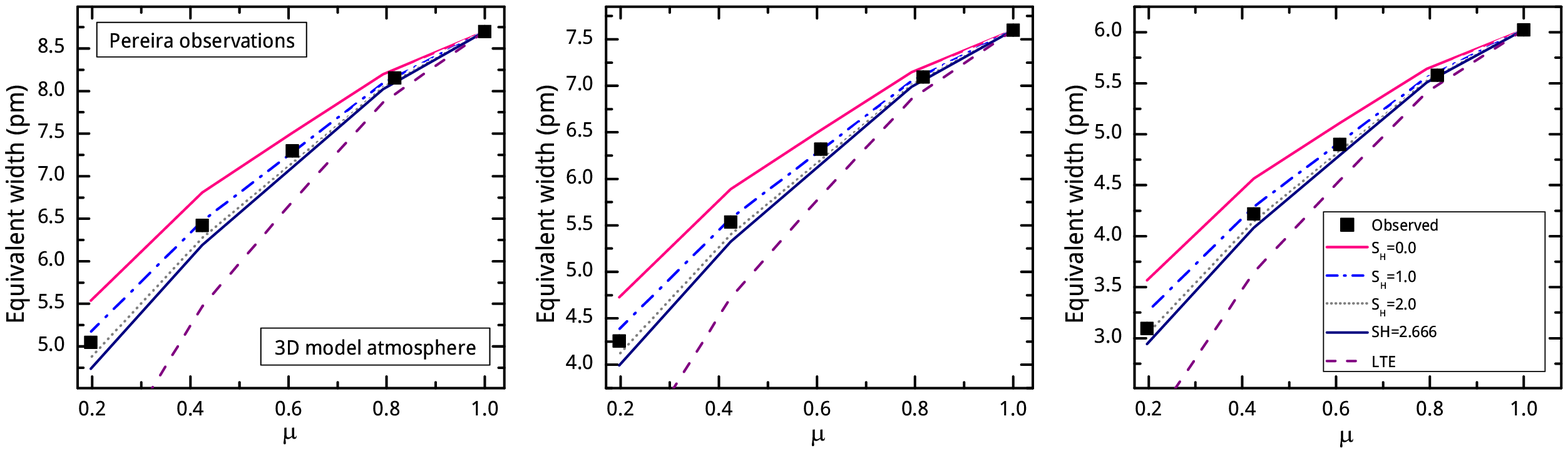}}
\mbox{\includegraphics*[bb=242 0 721 205,width=0.95\columnwidth]
{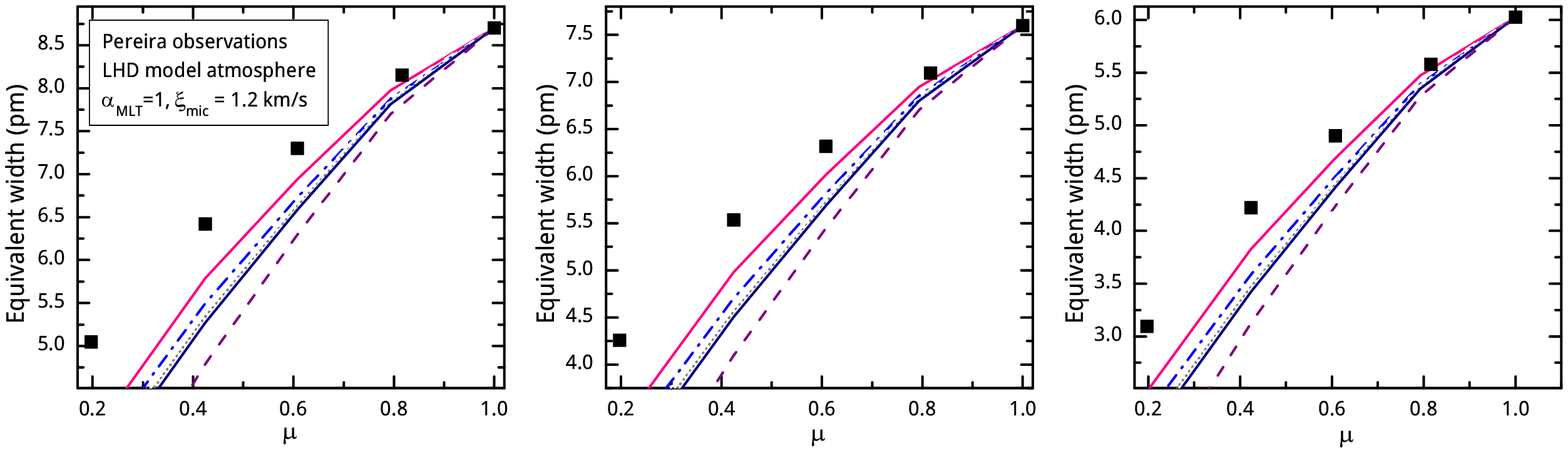}}
\caption{Center-to-limb variation of the equivalent width of the 777.4\,nm 
 (left panels) and 777.5\,nm (right panels ) components of the oxygen 
 triplet. The observed equivalent widths 
\citep[black squares,][]{Pereira2009a,Pereira2009b} are compared to
 the predictions of non-LTE line formation calculations for a 3D solar 
 \COBOLD\ model (top panels) and of a 1D hydrostatic (LHD) model assuming 
 a constant, isotropic  microturbulence $\ximic=1.2\pun{km/s}$ (bottom 
 panels). The theoretical results depend on the efficiency of collisions
 between oxygen and neutral hydrogen controlled by the parameter \SH.
\label{f:ximic}}
\end{figure}

As an example of the failure of 1D models with isotropic microturbulence,
we consider the comparison between the observed and predicted
center-to-limb variation of the 777\,nm solar oxygen triplet lines. 
These lines are partly saturated in the solar spectrum and
hence are sensitive to microturbulence. At the same time, the triplet lines
are prone to departures from local thermodynamic equilibrium (LTE), and
modeling their formation requires the solution of the statistical equilibrium 
equations for a multi-level oxygen model atom. 
The solution of this problem is rather sensitive to the strength of
collisions between oxygen and neutral hydrogen atoms. Unfortunately, the 
relevant collisional cross-sections are so far not known from laboratory
measurements or quantum-mechanical calculations. In this situation, the only
possibility is to empirically constrain the scaling factor \SH\ that 
quantifies the efficiency of inelastic collisions with neutral 
hydrogen atoms relative to the classical Drawin formula \citep{Drawin1969,
SteenbHolw1984} from the $\mu$-dependence of the line profiles or their 
equivalent widths. This procedure is illustrated in  Fig.\,\ref{f:ximic}.

From the 3D solar model, where the hydrodynamical velocity field
automatically accounts for an anisotropic, depth-dependent microturbulence,
the best fit to the observations is obtained with \SH\ between 1.0 and 1.6
(upper panels). In contrast, the 1D LHD model with constant, isotropic 
microturbulence would require a \emph{negative} value of \SH\ to reproduce 
the observations (lower panels), i.e.\ unphysical (negative) collisional 
cross-sections. A different choice of \ximic\ does not improve the 
situation. It is the assumption of a $\mu$-independent microturbulence 
that leads to the failure of the 1D model to reproduce the observed
center-to-limb variation of the solar oxygen triplet lines. Further
details can be found in \citet{Steffen+al15}.

\subsection{Implications for stellar abundance work}

The rudimentary description of the photospheric velocity field in
the 1D models, together with the ignorance of horizontal temperature 
and pressure fluctuations, implies basic limitations of the precision 
of the standard abundance analysis performed with 1D models. In the 
following, we give a quantitative example.

Imagine you have a noiseless stellar spectrum from which you can
measure the equivalent widths of a selection of iron lines with 
infinite precision; the atomic parameters of the lines are 
perfectly known. Moreover, the stellar parameters of the observed 
object (\Teff, \logg, [M/H]) have already been determined independently 
of spectroscopy to high precision. It is also known that all iron lines
form in strict LTE. Now the question is how well the known iron abundance 
can be reproduced from a spectroscopic analysis of the stellar spectrum 
using a 1D atmosphere model. 

The situation described above can be investigated by identifying the 
synthetic spectrum of a 3D hydrodynamical model with the observation,
and analyzing this spectrum in terms of the corresponding 1D model.
For the example shown below, we have constructed the 1D model by
averaging the 3D model on surfaces of constant Rosseland optical depth,
ensuring that the temperature structure of the resulting 1D model 
(= \xtmean{3D} model) matches exactly the mean temperature structure 
of the ``observed'' stellar photosphere. The only deficiencies of the 
1D model are its inability to account for horizontal fluctuations and 
its simplistic microturbulence velocity field.

\begin{figure}
\begin{center}
\mbox{\includegraphics*[bb=14 28 580 360,width=\columnwidth]
{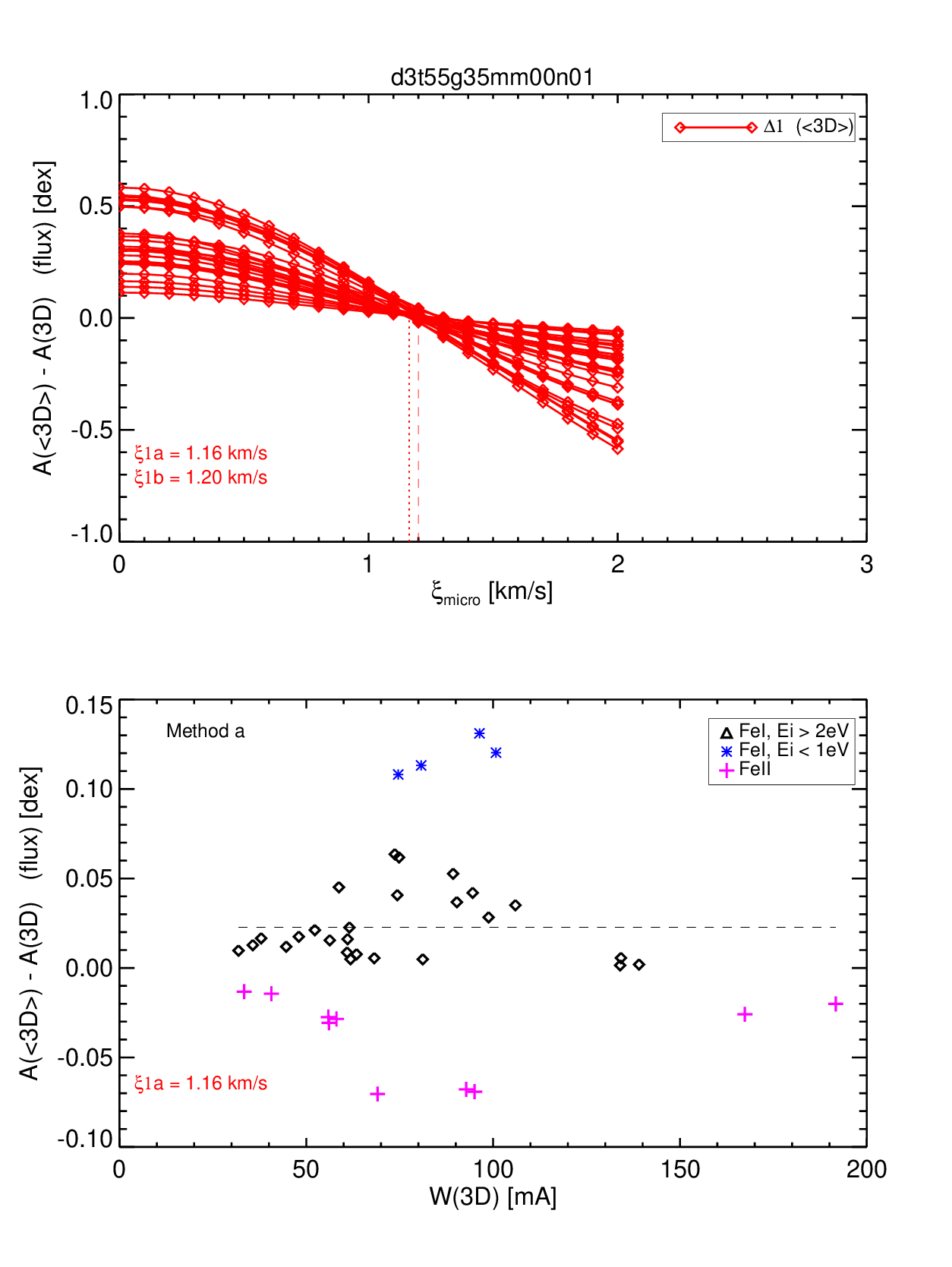}}
\mbox{\includegraphics*[bb=14 410 580 750,width=\columnwidth]
{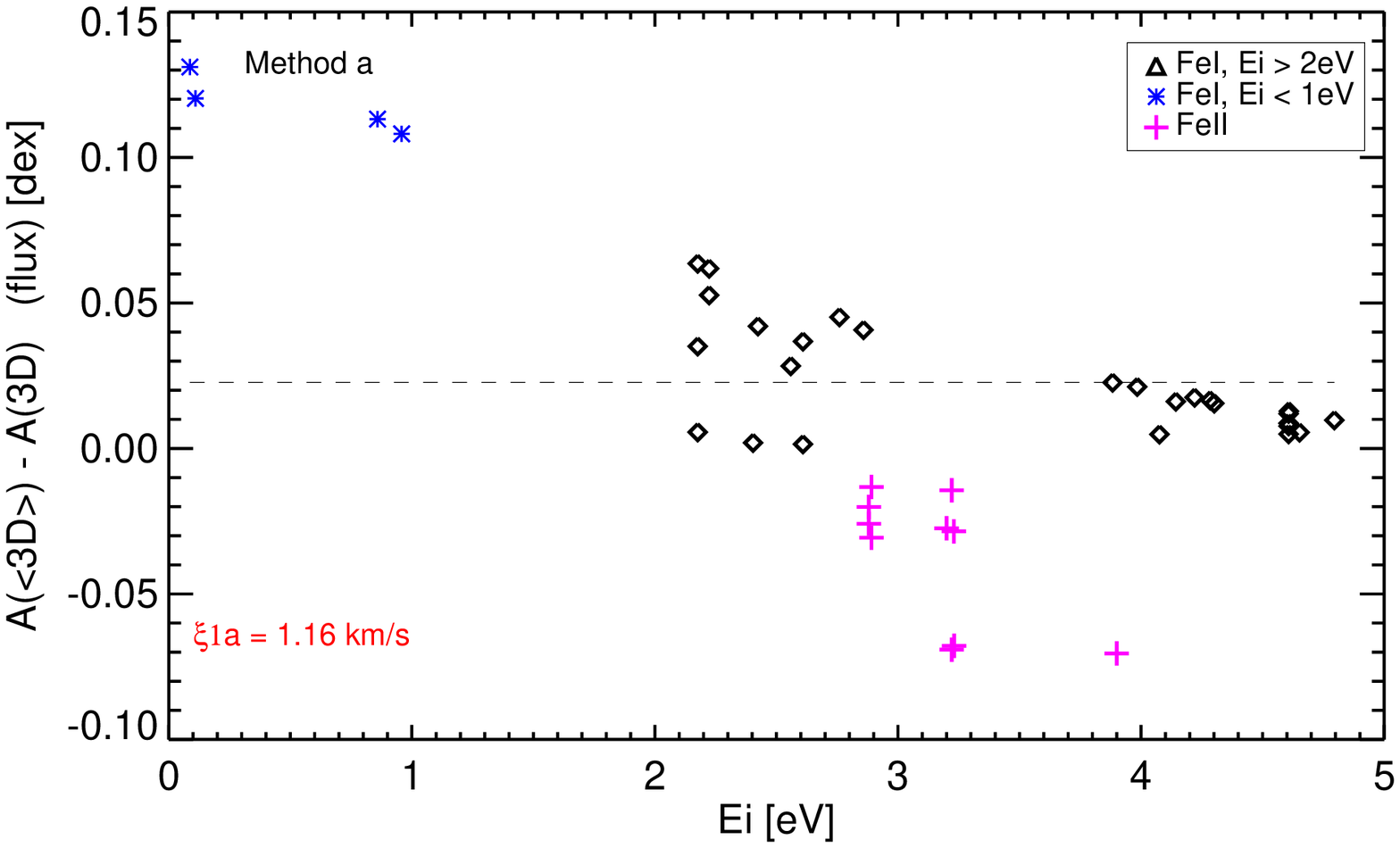}}
\end{center}
\caption{\xtmean{3D}\,$-$\,3D abundance differences for a set of 
\ion{Fe}{i} and \ion{Fe}{ii} lines versus equivalent width $W$ (\emph{top}) 
and lower level excitation potential $E$ (\emph{bottom}). \xtmean{3D} 
and 3D model have identical stellar parameters (\Teff\ = $5430$\,K, 
$\log g=3.5$, [M/H]=0). In the \xtmean{3D} model, \ximic\ was adjusted to 
remove the trend of abundance with $W$ for \ion{Fe}{i} lines with excitation 
potential $E>2$\,eV (black diamonds). % Other symbols see legend.
\label{f:micro_diag}}
\end{figure}

Figure \ref{f:micro_diag} illustrates this experiment for a subgiant with
\Teff\,=\,5500\,K, \logg\,=\,3.5, and solar metallicity. The top panel
shows the diagnostic plot from which the microturbulence was determined. 
Following common practice, \ximic\ was determined from \ion{Fe}{i} lines 
with excitation potential $E>2$\,eV (black diamonds) such that the trend
in the 1D abundance (difference) with equivalent width is removed. 
Low excitation lines of \ion{Fe}{i} and \ion{Fe}{ii} lines are shown for 
further information. 
The average iron abundance of the microturbulence lines (dashed horizontal 
line) comes out close to the abundance put into the 3D synthesis. 
However, it is remarkable that we are left with an overall scatter of 
$>0.2$\,dex peak-to-peak if all lines are included, even though 
\ximic\ has been properly adjusted, the 1D model has the correct \Teff\ 
and \logg, and the equivalent widths and $gf$-values of all lines are 
perfectly known. The residual abundance scatter 
must be a consequence of the imperfect representation of the photospheric 
velocity field by a constant microturbulence, and/or the 
inability of the 1D model to account for the presence of photospheric 
temperature fluctuations. 

We further note a systematic offset between \ion{Fe}{i} and
\ion{Fe}{ii} lines. In a classical spectroscopic analysis, this would be
compensated by increasing \logg\ of the 1D model to achieve ionization 
equilibrium. At the same time, there is a systematic trend with excitation 
potential $E$ (lower panel of Fig.\,\ref{f:micro_diag}), which would be 
eliminated by decreasing \Teff. This example demonstrates that we 
must not be surprised if the spectroscopic analysis based on 1D model 
atmospheres yields stellar parameters that are inconsistent with the 
true physical parameters \Teff\ and \logg\, (benchmark stars). 
3D Effects of similar magnitude have been shown to exist for an 
F-dwarf at \Teff$\,=\,$6300$\,K, \logg$\,=\,$4.5$, [M/H]=0 by 
\citet{Ludwig2014}.

\acknowledgements H.G.L. acknowledges financial support by the
Sonderforschungsbereich SFB 881 ``The Milky Way System'' (subproject A4) of
the German Research Foundation (DFG). We thank Dainius Prakapavi{\v c}ius for
preparing Fig.\,\ref{f:ximic}, and Jonas Klevas for extending the time series
of several 3D models.

% Use this code if you wish to generate your bibliography with BibTeX;
% please replace first the string "an-demo" below with the name(s) of
% the BibTeX data base(s) you want to use.
% The resulting bibliography-output (the contents of the .bbl file)
% must be pasted into this file before submission.
% 
\bibliographystyle{an}
\bibliography{ludwig}
% 
% Replace the following example bibliography with your references
% before submission:

\end{document}